\documentclass{PoS}

\title{Nucleon sea and the five-quark components}

\ShortTitle{Nucleon sea and the five-quark components}

\author{\speaker{Jen-Chieh Peng}\\
Department of Physics, University of Illinois at Urbana-Champaign, Urbana, IL 61801, USA\\
E-mail: \email{jcpeng@illinois.edu}}

\author{Wen-Chen Chang\\
Institute of Physics, Academia Sinica, Taipei 11529, Taiwan\\
E-mail: \email{changwc@phys.sinica.edu.tw}}

\abstract{
We generalize the approach of Brodsky {\it et al.} for the
intrinsic charm quark distribution in the nucleons to the light-quark
sector involving intrinsic $\bar u, \bar d , s$ and $\bar s$ sea
quarks. We compare the calculations with the existing $\bar d - \bar
u$, $s + \bar s$, and $\bar u + \bar d - s -\bar s$ data. The good
agreement between the theory and the data is interpreted as
evidence for the existence of the intrinsic light-quark sea in the
nucleons. The probabilities for the $|uudu\bar{u}\rangle$,
$|uudd\bar{d}\rangle$ and $|uuds\bar{s}\rangle$ Fock states are also 
extracted.
}

\FullConference{Sixth International Conference on Quarks and Nuclear
Physics\\
April 16-20, 2012\\
Ecole Polytechnique, Palaiseau,  Paris}

\begin{document}

\section{Introduction}

The origin of sea quarks of the nucleons remains a subject of intense
interest in hadron physics.
The possible existence of a significant $u u d c \bar c$ five-quark
Fock component in the proton was proposed some time ago by Brodsky,
Hoyer, Peterson, and Sakai (BHPS)~\cite{brodsky80} to explain the
unexpectedly large production rates of charmed hadrons at large
forward $x_F$ region. 
The intrinsic charm originating from the
five-quark Fock state is to be distinguished from the ``extrinsic"
charm produced in the splitting of gluons into $c \bar c$ pairs, which
is well described by QCD. The extrinsic charm has a ``sea-like"
characteristics with large magnitude only at the small $x$ region. In
contrast, the intrinsic charm is ``valence-like" with a distribution
peaking at larger $x$. The presence of the intrinsic charm component
can lead to a sizable charm production at the forward rapidity ($x_F$)
region.

The CTEQ collaboration~\cite{pumplin} has examined all relevant
hard-scattering
data and concluded that the data are consistent with a broad range of
the intrinsic charm magnitude, ranging from null to 2-3 times larger than
the estimate by the BHPS model. This suggests that more precise experimental
measurements are needed for determining the magnitude of the intrinsic
charm component.

In an attempt to further study the role of five-quark Fock states for
intrinsic quark distributions in the nucleons, we have extended the
BHPS model to the light quark sector and compared the predictions with
the experimental data. The BHPS model predicts the probability for the
$u u d Q \bar Q$ five-quark Fock state to be approximately
proportional to $1/m_Q^2$, where $m_Q$ is the mass of the quark
$Q$~\cite{brodsky80}. Therefore, the light five-quark states $u u d u
\bar u$, $u u d d \bar d$ and $u u d s \bar s$ are expected to have 
significantly larger
probabilities than the $u u d c \bar c$ state. This suggests that the
light quark sector could potentially provide more clear evidence for
the presence of the five-quark Fock states, allowing 
predictions of the BHPS model, such as the shape of the antiquark $x$
distributions originating from the five-quark configuration, to be
tested.

To search for evidence of the
intrinsic five-quark Fock states, it is essential to separate the
contributions of the intrinsic quark and the extrinsic
one. Fortunately, there exist some experimental observables which are
free from the contributions of the extrinsic quarks. As discussed
below, the $\bar d - \bar u$ and the $\bar u + \bar d - s - \bar s$
are examples of quantities independent of the contributions from
extrinsic quarks.
The $x$ distribution of $\bar d - \bar u$ has been
measured in Drell-Yan experiments~\cite{na51,e866}. A recent measurement
of $s + \bar s$ in a semi-inclusive deep-inelastic scattering (DIS)
experiment~\cite{hermes} also allowed the determination of the $x$
distribution of $\bar u + \bar d - s - \bar s$. In this paper, we
compare these data with the calculations based on the intrinsic
five-quark Fock states. The qualitative agreement between the data and
the calculations provides evidence for the existence of the intrinsic
light-quark sea in the nucleons~\cite{chang1}. We also show how 
the probabilities
of various fine-quark states can be determined~\cite{chang2}.

\section{Momentum distributions of five-quark states}

For a $|u u d Q \bar Q\rangle$ proton Fock state, the probability for
quark $i$ to carry a momentum fraction $x_i$ is given in the BHPS
model~\cite{brodsky80} as
\begin{equation}
P(x_1, ...,x_5)=N_5\delta(1-\sum_{i=1}^5x_i)[m_p^2-\sum_{i=1}^5\frac{m_i^2}{x_i}
]^{-2},
\label{eq:prob5q_a}
\end{equation}
\noindent where the delta function ensures momentum
conservation. $N_5$ is the normalization factor, 
and $m_i$ is the mass of quark $i$. In the limit of $m_{4,5} >>
m_p, m_{1,2,3}$, where $m_p$ is the proton mass, Eq.~\ref{eq:prob5q_a}
becomes
\begin{equation}
P(x_1, ...,x_5)=\tilde{N}_5\frac{x_4^2x_5^2}{(x_4+x_5)^2} \delta(1-\sum_{i=1}^5
x_i),
\label{eq:prob5q_b}
\end{equation}
\noindent where $\tilde{N}_5 = N_5/m_{4,5}^4$. Eq.~\ref{eq:prob5q_b}
can be readily integrated over $x_1$, $x_2$, $x_3$ and $x_4$, and the
heavy-quark $x$ distribution~\cite{brodsky80} is:
\begin{eqnarray}
P(x_5)=\frac{1}{2} \tilde{N}_5 x_5^2[\frac{1}{3} (1-x_5)
(1+10x_5+x_5^2)-2x_5(1+x_5)\ln (1/x_5)].
\label{eq:prob5q_d}
\end{eqnarray}
\noindent One can integrate Eq.~\ref{eq:prob5q_d} over $x_5$ and
obtain the result ${\cal P}^{c \bar c}_5 = \tilde{N}_5/3600$, where
${\cal P}^{c \bar c}_5$ is the probability for the $|u u d c \bar
c\rangle$ five-quark Fock state. An estimate of the magnitude of
${\cal P}^{c \bar c}_5$ was given by Brodsky et al.~\cite{brodsky80}
as $\approx 0.01$, based on diffractive production of
$\Lambda_c$. This value is consistent with a bag-model
estimate~\cite{donoghue77}.

\begin{figure}
\hspace{3.5cm}
\includegraphics[width=0.55\textwidth]{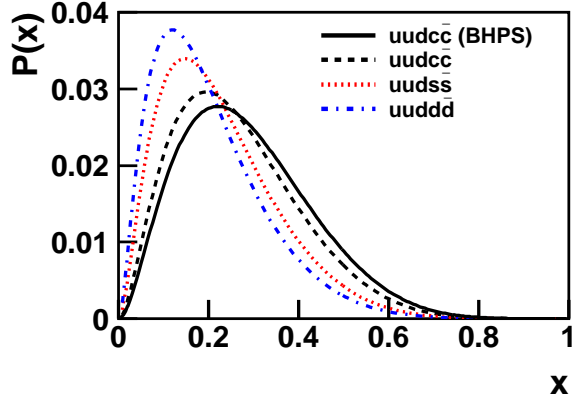}
\caption{The $x$ distributions of the intrinsic $\bar Q$ in the $u u d
Q \bar Q$ configuration of the proton from the BHPS
model. The solid curve is obtained using the
expression in Eq.~2.3 for $\bar c$. The other three
curves, corresponding to $\bar c$, $\bar s$, and $\bar d$ in the
five-quark configurations, are obtained by solving
Eq.~2.1 numerically. The same probability, 
${\cal P}^{Q \bar Q}_5= 0.01$, is 
used for the three different five-quark states.}
\label{fig_5q_c_s_d}
\end{figure}

The solid curve in Fig.~\ref{fig_5q_c_s_d} shows the $x$ distribution
for the charm quark ($P(x_5)$) using Eq.~\ref{eq:prob5q_d}, assuming
${\cal P}^{c \bar c}_5 = 0.01$. Since this analytical expression was
obtained for the limiting case of infinite charm-quark mass, it is of
interest to compare this result with calculations without such an
assumption. To this end, we have developed the algorithm to calculate
the quark distributions using Eq.~\ref{eq:prob5q_a} with Monte-Carlo
techniques. The five-quark configuration of $\{x_1,...,x_5\}$
satisfying the constraint of Eq.~\ref{eq:prob5q_a} is randomly
sampled. The probability distribution $P(x_i)$ can be obtained
numerically with an accumulation of sufficient statistics. We first
verified that the Monte-Carlo calculations in the limit of very heavy
charm quarks reproduce the analytical result for $P(x_5)$ in
Eq.~\ref{eq:prob5q_d}. We then calculated $P(x_5)$ using $m_u = m_d =
0.3$ GeV, $m_c = 1.5$ GeV, and $m_p = 0.938$ GeV,
and the result is shown as the dashed curve in
Fig.~\ref{fig_5q_c_s_d}. The similarity between the solid and dashed
curves shows that the assumption adopted for deriving
Eq.~\ref{eq:prob5q_d} is adequate. It is important to note that the
Monte-Carlo technique allows us to calculate the quark $x$
distributions for other five-quark configurations when $Q$ is the
lighter $u$, $d$, or $s$ quark, for which one could no longer assume a
large mass.

As shown in Fig.~\ref{fig_5q_c_s_d}, we have
calculated the $x$ distributions of the $\bar s$ and $\bar d$ quarks
in the BHPS model for the $|u u d s \bar s\rangle$ and $|u u d d \bar
d\rangle$ configurations, respectively, using
Eq.~\ref{eq:prob5q_a}. The mass of the strange quark is chosen as 0.5
GeV. In order to
focus on the different shapes of the $x$ distributions, the same value
of ${\cal P}^{Q \bar Q}_5$ is assumed for various five-quark
states. Figure~\ref{fig_5q_c_s_d} shows that the $x$ distributions of
the intrinsic $\bar Q$ shift progressively to lower $x$ as the
mass of the quark $Q$ decreases. The challenge
is to identify proper experimental observables which allow a clear
separation of the intrinsic light quark component from the extrinsic
QCD component. As we discuss next, the quantities $\bar d(x) - \bar
u(x)$, $s(x) + \bar s(x)$, and $\bar u(x) + \bar d(x) - s(x) - \bar s(x)$ 
are suitable for studying the intrinsic light-quark components of the proton.

\section{Extraction of various five-quark components}

\begin{figure}
\hspace{3.5cm}
\includegraphics[width=0.55\textwidth]{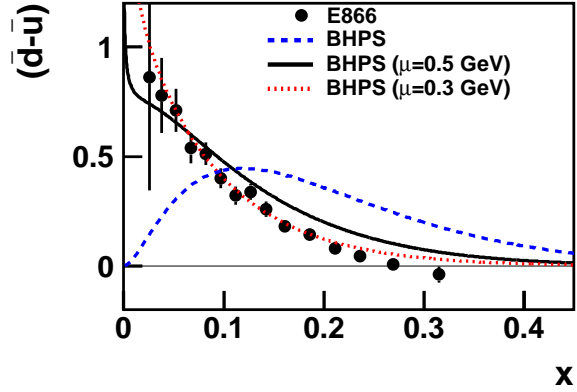}
\caption{Comparison of the $\bar d(x) - \bar u(x)$ data with the
calculations based on the BHPS model. The dashed curve corresponds to
the calculation using Eq.~2.1, and the solid and dotted curves are
obtained by evolving the BHPS result to $Q^2 = 54.0$ GeV$^2$ using
$\mu = 0.5$ GeV and $\mu = 0.3$ GeV, respectively.}
\label{fig_dbar-ubar}
\end{figure}

To compare the experimental data with the prediction based on the
intrinsic five-quark Fock state, it is necessary to separate the
contributions of the intrinsic sea quark and the extrinsic one. The
$\bar d(x) - \bar u(x)$ is an example of quantities which are free
from the contributions of the extrinsic sea quarks, since the
perturbative $g \to Q \bar Q$ processes will generate $u \bar u$ and
$d \bar d$ pairs with equal probabilities and have no contribution to
this quantity. The $\bar d(x) - \bar u(x)$ data from the Fermilab
E866 Drell-Yan experiment at the $Q^2$ scale of 54 GeV$^2$~\cite{e866}
are shown in Fig.~\ref{fig_dbar-ubar}.

In the BHPS
model, the $\bar u$ and $\bar d$ are predicted to have the same
$x$-dependence if $m_u = m_d$. However, the probabilities of the $|u u
d d \bar d\rangle$ and $|u u d u \bar u\rangle$ configurations, ${\cal
P}^{d \bar d}_5$ and ${\cal P}^{u \bar u}_5$, are not known from the
BHPS model, and remain to be determined from the
experiments. Non-perturbative effects such as
Pauli-blocking~\cite{feynman} could lead to different probabilities
for the $|u u d d \bar d\rangle$ and $|u u d u \bar u\rangle$
configurations. Nevertheless the shape of the $\bar d(x) - \bar u(x)$
distribution shall be identical to those of $\bar d(x)$ and $\bar
u(x)$ in the BHPS model. Moreover, the normalization of $\bar d(x) -
\bar u(x)$ is known from the measurement of Fermilab E866 Drell-Yan
experiment~\cite{e866} as
\begin{equation}
\int^{1}_{0} (\bar d(x) - \bar u(x)) dx =
{\cal P}^{d \bar d}_5 - {\cal P}^{u \bar u}_5 = 0.118 \pm 0.012.
\label{eq:intdbarubar1}
\end{equation}
Figure~\ref{fig_dbar-ubar} shows the calculation of the $\bar d(x) -
\bar u(x)$ distribution (dashed curve) from the BHPS model, together
with the data. The $x$-dependence of the $\bar d(x) - \bar u(x)$ data
is not in good agreement with the calculation. It is important to note
that the $\bar d(x) - \bar u(x)$ data in Fig.~\ref{fig_dbar-ubar} were
obtained at a rather large $Q^2$ of 54 GeV$^2$~\cite{e866}. In
contrast, the relevant scale, $\mu^2$, for the five-quark Fock states
is expected to be much lower, around the confinement scale. This
suggests that the apparent discrepancy between the data and the BHPS
model calculation in Fig.~\ref{fig_dbar-ubar} could be partially due
to the scale dependence of $\bar d(x) - \bar u(x)$. We adopt the value
of $\mu = 0.5$ GeV, which was chosen by Gl\"{u}ck, Reya, and
Vogt~\cite{grv} in their attempt to generate gluon and quark
distributions in the so-called ``dynamical approach" starting with
only valence-like distributions at the initial $\mu^2$ scale and
relying on evolution to generate the distributions at higher $Q^2$. We
have evolved the predicted $\bar d(x) - \bar u(x)$ distribution from
$Q_0^2 = \mu^2 =0.25$ GeV$^2$ to $Q^2 = 54$ GeV$^2$. Since $\bar d(x)
- \bar u(x)$ is a flavor non-singlet parton distribution, its
evolution from $Q_0$ to $Q$ only depends on the values of $\bar d(x) -
\bar u(x)$ at $Q_0$, and is independent of any other parton
distributions. The solid curve in Fig.~\ref{fig_dbar-ubar} corresponds
to $\bar d(x) - \bar u(x)$ from the BHPS model evolved to $Q^2=$ 54
GeV$^2$. Significantly improved agreement with the data is now
obtained. This shows that the $Q^2$-evolution should be properly
taken into account. It is interesting to note that an excellent fit to the
data can be obtained if $\mu = 0.3$ GeV is chosen (dotted curve in
Fig.~\ref{fig_dbar-ubar}) rather than the more conventional value of
$\mu = 0.5$ GeV. We have also found good agreement between the HERMES
$\bar d(x) - \bar u(x)$ data at $Q^2 = 2.3$ GeV$^2$~\cite{hermes_sidis}
with calculation using the BHPS model.

\begin{figure}
\hspace{3.5cm}
\includegraphics[width=0.55\textwidth]{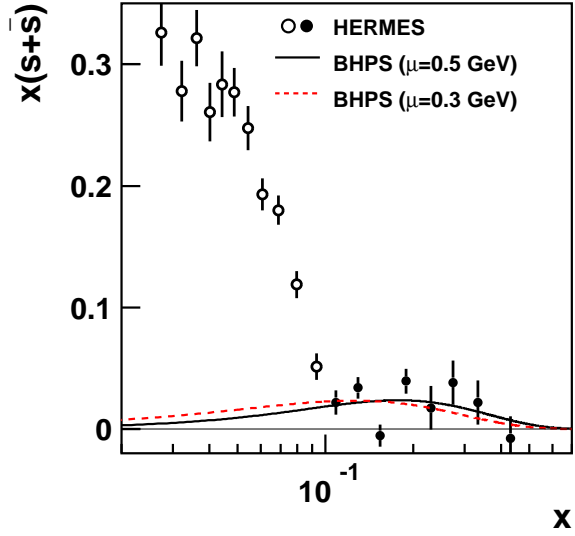}
\caption{Comparison of the HERMES $x(s(x) + \bar s(x))$ data with the
calculations based on the BHPS model. The solid and dashed curves are
obtained by evolving the BHPS result to $Q^2 = 2.5$ GeV$^2$ using $\mu
= 0.5$ GeV and $\mu = 0.3$ GeV, respectively. The normalizations of
the calculations are adjusted to fit the data at $x > 0.1$.}
\label{fig_ssbar}
\end{figure}

We now consider the extraction of the $|uuds \bar s\rangle$ five-quark
component from existing data. The HERMES collaboration reported the
determination of $x(s(x) + \bar s(x))$ over the range of $0.02 < x <
0.5$ at $Q^2 = 2.5$ GeV$^2$ from the measurement of charged kaon
production in semi-inclusive DIS reaction~\cite{hermes}. The HERMES 
data, shown in
Fig.~\ref{fig_ssbar}, exhibits an intriguing feature. A rapid fall-off
of the strange sea is observed as $x$ increases up to $x \sim 0.1$,
above which the data become relatively independent of $x$. The data
suggest the presence of two different components of the strange sea,
one of which dominates at small $x$ $(x<0.1)$ and the other at larger
$x$ $(x>0.1)$. This feature is consistent with the expectation that
the strange-quark sea consists of both the intrinsic and the extrinsic
components having dominant contributions at large and small $x$
regions, respectively. In Fig.~\ref{fig_dbar-ubar} we compare the 
data with calculations
using the BHPS model with $m_s=0.5$ GeV. The solid and dashed
curves are results of the BHPS model calculations evolved to $Q^2 =
2.5$ GeV$^2$ using $\mu = 0.5$ GeV and $\mu = 0.3$ GeV,
respectively. The normalizations are obtained by fitting only data
with $x>0.1$ (solid circles in Fig.~\ref{fig_ssbar}), following the
assumption that the extrinsic sea has negligible contribution relative
to the intrinsic sea in the valence region. Figure~\ref{fig_ssbar} 
shows that the
fits to the data are quite adequate, allowing the extraction of the
probability of the $|uuds \bar s\rangle$ state as
\begin{eqnarray}
{\cal P}^{s \bar s}_5 = 0.024~~(\mu = 0.5~\rm{GeV});~~~
{\cal P}^{s \bar s}_5 = 0.029~~(\mu = 0.3~\rm{GeV}).
\label{eq:ssbar_value}
\end{eqnarray}

\begin{figure}
\hspace{3.5cm}
\includegraphics[width=0.55\textwidth]{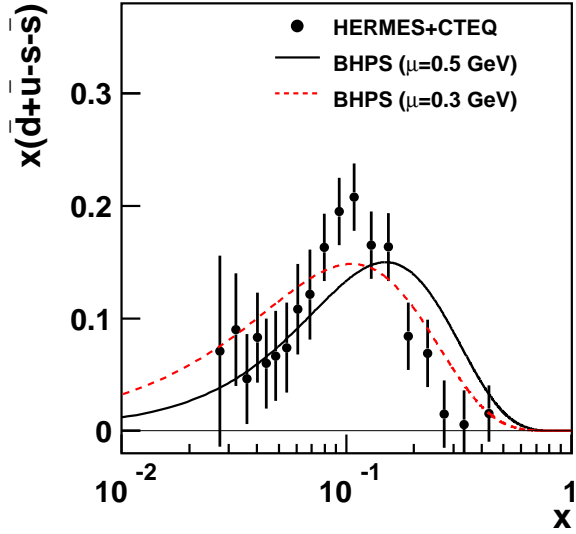}
\caption{Comparison of the $x(\bar d(x) + \bar u(x) - s(x) - \bar
s(x))$ data with the calculations based on the BHPS model. The values
of $x(s(x) + \bar s(x))$ are from the HERMES experiment~\cite{hermes},
and those of $x(\bar d(x) + \bar u(x))$ are obtained from the PDF set
CTEQ6.6~\cite{cteq}. The solid and dashed curves are
obtained by evolving the BHPS result to $Q^2 = 2.5$ GeV$^2$ using $\mu
= 0.5$ GeV and $\mu = 0.3$ GeV, respectively. The normalization of
the calculations are adjusted to fit the data.}
\label{fig_sbar-dubar}
\end{figure}

We consider next the quantity $\bar u(x) + \bar d(x) - s(x) - \bar
s(x)$. Combining the HERMES data on $x(s(x)+ \bar s(x))$ with the
$x(\bar d(x) + \bar u(x))$ distributions determined by the CTEQ group
(CTEQ6.6)~\cite{cteq}, the quantity $x(\bar u(x) + \bar d(x) - s(x) -
\bar s(x))$ can be obtained and is shown in
Fig.~\ref{fig_sbar-dubar}. This approach for determining $x(\bar u(x)
+ \bar d(x) - s(x) - \bar s(x))$ is identical to that used by Chen,
Cao, and Signal in their study of strange quark sea in the
meson-cloud model~\cite{signal}.

An important property of $\bar u + \bar d - s - \bar s$ is that the
contribution from the extrinsic sea vanishes, just like the case for
$\bar d - \bar u$. Therefore, this quantity is only sensitive to the
intrinsic sea and can be compared with the calculation of the
intrinsic sea in the BHPS model. We have
\begin{eqnarray}
\bar u(x) + \bar d(x) - s(x) - \bar s(x) = P^{u \bar u}(x_{\bar u}) +
P^{d \bar d}(x_{\bar d}) - 2 P^{s \bar s}(x_{\bar s}).
\label{eq:udssbar_p5}
\end{eqnarray}
\noindent where $P^{Q \bar Q}(x_{\bar Q})$ refers to the $x$-distribution of 
$\bar Q$ in the $|u u d Q \bar Q\rangle$ state. 
We can now compare the $x(\bar u(x) + \bar d(x) - s(x) -
\bar s(x))$ data with the calculation using the BHPS model. Since
$\bar u + \bar d - s - \bar s$ is a flavor non-singlet quantity, we
can readily evolve the BHPS prediction to $Q^2 =2.5$ GeV$^2$ using
$\mu = 0.5$ GeV and the result is shown as the solid curve in
Fig.~\ref{fig_sbar-dubar}. It is interesting to note that a better fit
to the data can again be obtained with $\mu = 0.3$ GeV, shown as the
dashed curve in Fig.~\ref{fig_sbar-dubar}.

From the comparison between the data and the BHPS calculations shown
in Figs.~\ref{fig_dbar-ubar}-\ref{fig_sbar-dubar}, we can determine
the probabilities for the $|u u d u \bar u\rangle$, $|u u d d \bar
d\rangle$, and $|u u d s \bar s\rangle$ configurations as follows:
\begin{eqnarray}
{\cal P}^{u \bar u}_5 = 0.122;~~{\cal P}^{d \bar d}_5 =
0.240;~~{\cal P}^{s \bar s}_5 = 0.024~~~ (\mu = 0.5~\rm{GeV})
\label{eq:uds_value_a}
\end{eqnarray}
\noindent or
\begin{eqnarray}
{\cal P}^{u \bar u}_5 = 0.162;~~{\cal P}^{d \bar d}_5 =
0.280;~~{\cal P}^{s \bar s}_5 = 0.029~~~ (\mu = 0.3~\rm{GeV})
\label{eq:uds_value_b}
\end{eqnarray}
\noindent depending on the value of the initial scale $\mu$. It is
remarkable that the $\bar d(x) - \bar u(x)$, the $s(x) + \bar s(x)$,
and the $\bar d(x) + \bar u(x) - s(x) - \bar s(x)$ data not only allow
us to check the predicted $x$-dependence of the five-quark Fock
states, but also provide a determination of the probabilities for
these states.

Equation~\ref{eq:uds_value_a} shows that the combined probability 
for proton to be in
the $|uud Q \bar Q\rangle$ states is around 40\%. It is worth noting that
an earlier analysis of the $\bar d - \bar u$ data in the meson cloud model
concluded that proton has $\sim$60\% probability to be in the three-quark
bare-nucleon state~\cite{szczurek}, in qualitative agreement with the
finding of this study.  A significant outcome of the present work is
the extraction of the $|uud s \bar s\rangle$ component, which is
related to the kaon-hyperon states in the meson cloud model. It is also
worth noting that the $|uud Q \bar Q\rangle$ states
have the same contribution to the proton's magnetic moment as the
$|uud\rangle$ three-quark state, since $Q$ and $\bar Q$ in the
$|uud Q \bar Q\rangle$ states have no net magnetic moment. Therefore,
the good description of the nucleon's magnetic moment by the constituent
quark model is preserved even with the inclusion of a sizable five-quark
components.

We note that the probability for the $|uud s \bar s\rangle$ state is
smaller than those of the $|uud u \bar u\rangle$ and the $|uud d \bar
d\rangle$ states. This is consistent with the expectation that the
probability for the $|uud Q \bar Q\rangle$ five-quark state is roughly
proportional to $1/m^2_Q$~\cite{brodsky80,franz00}. One can now
estimate that the probability
for the intrinsic charm from the $|uud c \bar c\rangle$ Fock state,
${\cal P}^{c \bar c}_5$ to be roughly $0.1 {\cal P}^{s \bar s}_5\sim 0.003$.
This shows that the probability of intrinsic charm
could be smaller than the earlier expectation~\cite{brodsky80}. Moreover,
the $Q^2$-evolution would shift the intrinsic-charm distribution to smaller
$x$, suggesting that the most promising region to search
for evidence of intrinsic charm could be at the somewhat lower $x$
region $(0.1 < x < 0.4)$, rather than the largest $x$ region explored
by previous experiments. 

\section{Conclusion}

In conclusion, we have generalized the existing BHPS model to the
light-quark sector and compared the calculation with the $\bar d -
\bar u$, $s + \bar s$, and $\bar u + \bar d - s - \bar s$ data. The
qualitative agreement between the data and the calculations provides
strong support for the existence of the intrinsic $u$, $d$ and $s$
quark sea and the adequacy of the BHPS model. This analysis also led
to the determination of the probabilities for the five-quark Fock
states for the proton involving light quarks only. This result could
guide future experimental searches for the intrinsic $c$ quark sea or
even the intrinsic $b$ quark sea~\cite{ma}, which could be relevant for
the production of Higgs boson at LHC energies~\cite{brodsky06}.
This analysis could also be readily extended to the hyperon and meson
sectors. The connections between the BHPS model, the meson-cloud 
model~\cite{tony}, the multi-quark
models~\cite{zhang,bourrely}, and the lattice QCD 
calculations~\cite{kfliu} should also be investigated.

\end{document}